\documentclass[letterpaper,twocolumn,showpacs,prl,superscriptaddress,floatfix]{revtex4}
\usepackage[german,swedish,english]{babel}
\usepackage[dvips]{graphicx}
\usepackage{amssymb}
\usepackage{amsmath}
\usepackage{color}

\begin{document}

\date{\today}

\title{Roton-like mode in solid $^4$He}

\author{E. Blackburn}
\affiliation{Physics Department, University of California, San Diego, La Jolla, California 92093, USA\\}
\author{S. K. Sinha}
\affiliation{Physics Department, University of California, San Diego, La Jolla, California 92093, USA\\}
\author{C. Broholm}
\affiliation{Department of Physics and Astronomy, The Johns Hopkins University, Baltimore, Maryland 21218, USA\\}
\author{J. R. D. Copley}
\affiliation{National Institute of Standards and Technology, Gaithersburg, Maryland 20899, USA\\}
\author{R. W. Erwin}
\affiliation{National Institute of Standards and Technology, Gaithersburg, Maryland 20899, USA\\}
\author{J. M. Goodkind}
\affiliation{Physics Department, University of California, San Diego, La Jolla, California 92093, USA\\}

\begin{abstract}
Recent theoretical and experimental work on solid $^4$He has focused on vacancies, or other lattice defects and whether these can form a Bose-Einstein condensate within the solid.  The earliest suggestion that this could happen was based on the conjecture that the ground state of the solid at $T$ = 0 K would include vacancies.  Although no direct evidence for zero point vacancies has been presented, a variety of experimental observations have been interpreted as being due to thermally activated vacancies. Here we directly probe the excitation spectrum of hcp solid $^4$He using inelastic neutron scattering.  We identify a branch of delocalized excitations, with both longitudinal and transverse dispersion that coexist with acoustic phonons. While the energy gap is larger and the characteristic wavevector is shifted to coincide with the position of the (100) Bragg peak, the longitudinal effective mass of this mode is within 15\% of that for rotons in superfluid $^4$He. The dispersion relation intersects the longitudinal acoustic phonon half way to the zone boundary, and this distinguishes the mode from a conventional optic phonon. Our results provide direct evidence for coherent delocalized non-phonon excitations in solid helium, which we identify as delocalized vacancy modes. 
\end{abstract}


\maketitle

\section{Introduction}
Theoretical speculation about possible Bose-Einstein condensation and superfluidity in solid $^4$He began almost 40 years ago \cite{AL69,IS75,LF73,Leggett70}.  Early experimental searches, using macroscopic techniques, found no evidence for the `supersolid' state \cite{BPR84,Suzuki73}.   However, evidence for some type of phase transition in impure or strained crystals was found using ultrasonic propagation \cite{HBG97,Goodkind02}.  Recent observations of a decrease in the moment of inertia of the solid, in impure or strained crystals, at temperatures below about 200 mK were interpreted as evidence for a supersolid \cite{KC04a,RR06,SICM+06}.  However, an  attempt to force direct flow through small channels was unsuccessful \cite{DB06}, and deep inelastic neutron scattering finds no evidence for a zero-momentum condensate fraction \cite{AMKD07,DPAK+07}.

As in the case of superfluid helium, the excitations in the solid could provide microscopic evidence for a possible Bose condensate or supersolid behavior if present.  Indirect measurements of the excitations have been made using NMR, ultrasound, pressure, and x-ray and neutron diffraction.  The results are generally interpreted in terms of vacancy excitation \cite{ARS82,MHMH74,FGS89,LG90,RSJF+97,BGSH+07}.   Vacancies are thermally activated and the activation energy increases roughly proportionally to the melting temperature over a wide range of densities \cite{MHMH74,LG90}.  However, the specific heat inferred from these activation energies is greater than the total measured specific heat if they are assumed to be localized, and is also too large if the vacancies behave like an ideal Bose gas \cite{BG94}.  A possible small contribution of these excitations to the pressure has been found \cite{RSJF+97}.  There is no direct experimental evidence for of zero-point vacancies in solid helium, and quantum Monte-Carlo calculations of perfect crystals do not find such vacancies \cite{CC06,BPS06}.

Here we report inelastic neutron scattering measurements on low density solid $^4$He crystals.  The measurements reveal portions of the spectrum that were not previously observable and provide higher resolution than earlier experiments \cite{MKLN+68}.  We find no evidence for a phase transition within the solid state but we do provide evidence for dispersive non-phonon excitations.  We identify these excitations as waves of delocalized vacancies and estimate their contribution to the specific heat close to 1 K.   Certain similarities between parts of the vacancy wave excitations and the roton spectrum in the superfluid are suggestive of strong links between the solid and liquid phases.

\section{Methods}
The excitation spectra were measured using the NG4 disk chopper time-of-flight neutron spectrometer DCS at the NIST Center for Neutron Research, Gaithersburg, MD \cite{CC03}.  The DCS uses choppers to create pulses of monochromatic neutrons.   Energy transfers on scattering are determined from arrival times in detectors that range in scattering angle from -30$^{\circ}$ to 140$^{\circ}$ in the horizontal plane and subtend a solid angle of $\approx$ 0.71 sr at the sample.  Unless otherwise mentioned, we operated in the 'low resolution' setup \cite{CC03} which gives a resolution of $\approx$ 0.2 meV full-width half-maximum (FWHM) at an incident wavelength of using 3.7 \AA~wavelength (energy 6.0 meV).

The $^4$He crystals were grown in a stainless steel cylindrical sample cell with the axis along the vertical.  The inside diameter of the cell was 20.5 mm and the outside diameter was 22 mm.   The neutron beam was restricted to a height of 25 mm (smaller than the sample height) using a cadmium mask.  The total volume of the cell was approximately 30 cm$^3$.  The lower flange of the sample chamber was machined out of Cu and included a ring of sintered Cu powder to enhance thermal contact to the He sample.  Contact between this Cu flange and the mixing chamber of the dilution refrigerator was made with an annealed copper ribbon 1.9 mm wide by 0.55 mm thick.  The upper flange was made of brass and provided a mechanical mount to the mixing chamber. X-cut quartz transducers (resonant frequency 9.77 MHz) were mounted at the centers of the top and bottom flanges to measure the acoustic velocity and attenuation through the helium along the vertical axis.  Two additional transducers were mounted in the top flange outside the neutron beam, to measure the acoustic properties in a horizontal direction.  The helium prepared for the samples contained $\approx$ 0.004 \% $^3He$ (40 ppm) at the start of the experiment but was diluted by a factor of 5 to 10 by the time the neutron measurements were made.  

Temperature was measured by a RuO$_2$ resistance thermometer attached to the mixing chamber. The base temperature reached by that thermometer was 55 mK. The ultrasonics were used to monitor crystal growth, as an internal thermometer of the solid helium, and as an indicator of crystal quality.  The large size of this sample relative ($\approx$ 30 cm$^3$) to that of previous acoustic measurements ($\approx$ 1 cm$^3$ e.g. Ref.~\onlinecite{HBG97}) and the long acoustic path (44.5 mm versus 8.43 mm) made it difficult to grow perfect crystals through the entire acoustic path.  In perfect crystals, the temperature dependence of the velocity is given by $\frac{v(T) - v(0\,\mathrm{K})}{v(0\,\mathrm{K})} = -$ 6.3 $\times$10$^{-4}$T$^4$ \cite{LG90}, and the acoustic attenuation, $\alpha$, varies as $\alpha \propto T^{3/2} \exp(-\Delta/T)$, where $\Delta$ is associated with the vacancy activation energy.  In strained crystals, the attenuation shows a broad maximum around 0.5 K and the fractional change in the velocity from $T \ll$ 1 K to $T$ = 1 K is much more than 6.3$\times$10$^{-4}$ and departs strongly from the $T^4$ dependence.  In the first crystal measured in this work, the fractional velocity change between 1 K and base temperature was 3.4$\times$10$^{-3}$, indicating significant strain.   The $T$ dependences of $v$ and $\alpha$ were also characteristic of crystals containing non-zero dislocation densities \cite{HBG97}.  This crystal also displayed a small acoustic anomaly like that found in Ref.~\onlinecite{HBG97}, confirming that measurements were made at temperatures below the acoustic anomaly at 200 mK in solid hcp $^4$He.  The crystals examined in the work described here were grown from the liquid at nearly constant pressure and temperature, not by the blocked capillary technique. 

The first sample (Crystal 1) measured was grown with the mixing chamber at 1 K.  The sample grew as a single crystal with the [001] axis vertical to within $< 1^{\circ}$.  The $a$ lattice parameter was 3.68 \AA.  The absolute value of the acoustic velocity (450 ms$^{-1}$) was also consistent with this orientation \cite{Greywall71}.  Using the spectrometer in a white-beam (Laue) mode all peaks observed were associated with the $c$-axis oriented single crystal.  The second sample was grown with the mixing chamber at 0.5 K and consisted of several large crystals, oriented in different directions.  The orientation of one of these crystals (Crystal 2) permitted acquisition of clean spectra of the longitudinal excitations about the (100) Bragg peak.  The temperature inside the sample chamber during growth of the crystals was higher than that of the mixing chamber since the heat leak through superfluid in the filling capillary is much higher than when the capillary is filled with solid helium.  Measurements with a thermometer inside a similar sample chamber indicate that the growth temperatures were approximately 1.2 K.  In both cases, the observed spectra showed no evidence of liquid helium in the path of the beam.   Spectra were also obtained for the superfluid at 24 bar, the empty cell at $T$ = 5 K and a vanadium standard was used to calibrate detector efficiencies. Unless otherwise stated all errors refer to one standard deviation.

\section{Results}
\begin{figure}[tbhp]
\begin{center}
\includegraphics[width=0.4\textwidth]{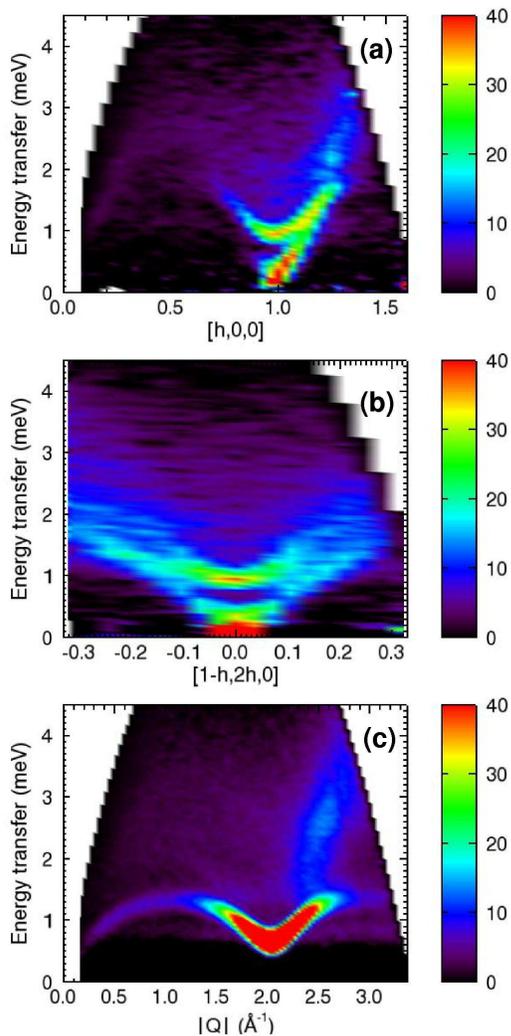}
\end{center}
\caption{(a) Cut along the [h,0,0] direction, integrating over the transverse direction by $\pm$ 0.01 reciprocal lattice units (r.l.u.).  (b) Cut of points (1-$h$,2$h$,0) (integrating over $\pm$ 0.01 r.l.u.).    These data were taken at base temperature, with an incident wavelength of 3.7 \AA.  (c) Superfluid data, taken at a pressure of 24 bar and $\approx$ 0.22 K with an incident wavelength of 3.49 \AA.  All data have been treated using empty cell and vanadium data as described.  The (100) Bragg peak has an absolute momentum transfer of 1.97 \AA $^{-1}$.}
\label{fig:overview}
\end{figure}

Figures \ref{fig:overview}a and \ref{fig:overview}b show scattering intensity contours in energy-wavevector space for wavevectors in the basal plane, both transverse and longitudinal relative to the (100) Bragg peak for Crystal 1.  The $a$ lattice parameter was 3.69 \AA.  A range of incident energies were used, to provide greater coverage in momentum space or better energy resolution.  For each incident energy, at least 21 different orientations of the crystal relative to the incident beam were used, covering a rotation of 42$^{\circ}$ about the axis perpendicular to the scattering plane.  For comparison Fig.~\ref{fig:overview}c shows the corresponding inelastic spectra for the superfluid at a pressure just below the freezing pressure (24 bar). The dominant new features in the longitudinal cut (Fig.~\ref{fig:overview}a) are:
\begin{itemize}
\item a quadratic dispersion curve centered at the (100) reciprocal lattice point.
\item the apparent intersection and subsequent disappearance of the two sharp longitudinal modes about half way to the zone boundary. 
\item a marked asymmetry in the intensity of the acoustic phonon which reduces the point group symmetry to triangular \cite{RSBT71,MKLN+68}.
\item a broad spectral feature that appears at about 2 meV, for wavevectors beyond the point of intersection mentioned above, and extends beyond the energy and wavevector range measured here.  
\end{itemize}

\begin{figure}[tbhp]
\begin{center}
\includegraphics[width=0.4\textwidth]{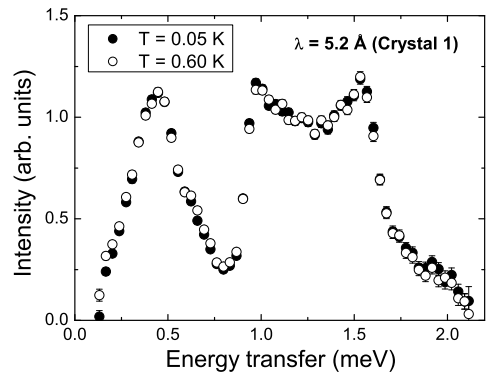}
\end{center}
\caption{Neutron weighted density of states integrating over all measured momentum transfers at base temperature (closed circles) and high temperature (open circles) for an incident neutron wavelength of 5.2 \AA.  Using the thermometer placed on the mixing chamber, base temperature is 0.055 K and high temperature is 0.6 K.}
\label{fig:tempdep}
\end{figure}

This latter broad feature is also seen in the superfluid (Fig.~\ref{fig:overview}c), starting at a similar absolute momentum transfer ($\approx$ 2.4 \AA $^{-1}$) and a slightly lower energy, corresponding to approximately twice the minimum energy gap in each case.  This feature is also seen in the normal fluid \cite{PAFS+01}.  In the solid, the high energy excitation displays hexagonal symmetry about the origin.
In Ref.~\cite{PAFS+01}, this scattering was identified as broad, multiphonon scattering.  Alternatively, it might also be associated with strongly interacting single-particle scattering, in which case it should connect with the free particle recoil scattering at larger momentum transfers.  Comparing Figs.~\ref{fig:overview}a and \ref{fig:overview}c, the spectra appear superficially to be very similar.  Glyde and Griffin \cite{GG90} have argued that, in the superfluid, the maxon and roton can be interpreted as single particle excitations, closely coupled to the existence of a superfluid condensate.

No differences in the excitation spectra are observed (Fig.~\ref{fig:tempdep}) above and below the temperature where both an acoustic anomaly and the onset of a decrease in moment of inertia were observed ($\approx$ 200 mK).  The longitudinal data were all confirmed on Crystal 2.    

Some high energy resolution data was also taken to look for (dispersionless) excitations with energies $\approx$ 1 K.  Such modes had been found using ultrasound \cite{LG90}.  We could not resolve any such excitations in measurements spanning approximately half the first Brillouin zone and the energy range from 0 meV to 1.2 meV.  The nominal energy resolution for the spectrometer in this configuration was 54 $\mathrm{\mu}$eV (FWHM) for an energy transfer of 0 meV.  We note that in Ref.~\onlinecite{LG90} this low activation energy was only seen in crystals grown below 1 K.  The growth temperature in the current work is estimated to be 1.2 K or slightly higher.

\section{Vacancy Excitations}

\begin{figure*}[t]
\begin{center}
\includegraphics[width=0.8\textwidth]{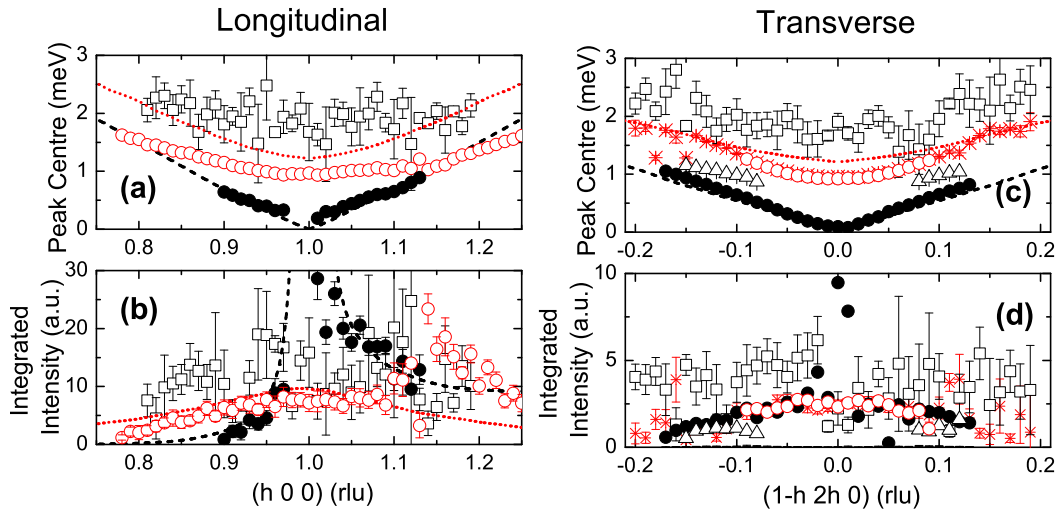}
\end{center}
\caption{Excitations in the longitudinal and transverse directions from the (100) Bragg peak, measured at 0.055 K with a neutron wavelength of 3.7 \AA~for the longitudinal cuts (a,b) and 5.2 \AA~for the transverse cuts (c,d).  The positions (a,c) and intensities (b,d) were obtained from Gaussian fitting.  The acoustic phonon branches are shown by filled circles, the vacancy mode by open circles, the optic branches by open squares, and in the transverse case there is an additional mode marked by open triangles.  The neutron wavelength was 3.7 \AA~for the longitudinal cuts (a,b) and 5.2 \AA~for the transverse cuts (c,d), apart from the optic mode in the transverse direction, which was measured at 3.7 \AA.  For comparison the vacancy mode at 3.7 \AA~is also shown (asterisks) in (b) and used to scale the intensity of the 3.7 \AA data in (d) to that at 5.2 \AA.  In the upper panels (a,c), the lines are the calculated dispersion curves for the acoustic (dashed) and optic (dotted) phonon branches from the self-consistent phonon calculations in Ref.~\onlinecite{GKW68}.  In (b), the intensities are calculated from the inelastic structure factor following Ref.~\onlinecite{GKW68} with an additional common scaling factor.}
\label{fig:spectrum}
\end{figure*}

We now examine the low energy excitations close to the (100) Bragg point more carefully.  Figure \ref{fig:spectrum} shows the dispersions of these excitations, as obtained by fitting Gaussians to the data.  The lowest energy mode is an acoustic phonon and we argue that the sharp intermediate energy mode is a collective vacancy excitation.  Minkiewicz {\it et al.}~\cite{MKLN+68} associated the intermediate energy mode with an optical phonon despite the discrepancy with the self-consistent phonon calculations of Gillis, Koehler and Werthamer \cite{GKW68}, to be discussed below.  The intensities for the longitudinal branches were calculated from the inelastic structure factor based on this self-consistent phonon theory \cite{GKW68,RSBT71}, and compare favorably to the data in Fig.~\ref{fig:spectrum} except at the intersection.  In particular, the asymmetry of the acoustic phonon branch is well reproduced.   

\begin{figure}[tbhp]
\begin{center}
\includegraphics[width=0.4\textwidth]{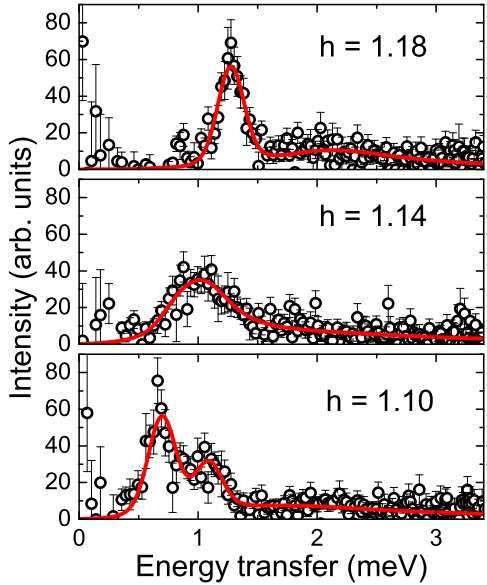}
\end{center}
\caption{Energy scans at ($h$00) integrating over a momentum transfer of $h$ $\pm$ 0.005 r.l.u.~in the longitudinal direction and $\pm$ 0.01 r.l.u.~in the transverse direction, for three selected values of H.  The wavelength used was 3.7 \AA, at base temperature.  The solid lines are fits using a model consisting of three damped harmonic oscillators convoluted with a Gaussian resolution function of 0.2 meV FWHM.  The positions of two of the oscillators are fixed using the phonon dispersions calculated by Gillis, Koehler and Werthamer \cite{GKW68}, and the third (middle) oscillator determines the position of the vacancy mode.  At h = 1.18 the lower and intermediate energy peaks have merged into a single peak and the spectrum can be described by two modes.}
\label{fig:intersect}
\end{figure}

We present two arguments for associating the intermediate energy mode with a vacancy excitation.  (i) This mode intersects with the longitudinal acoustic phonon branch.  On symmetry grounds, longitudinal optic and acoustic phonon branches in a hcp structure may not cross or be degenerate except at the zone boundary in the $c^*$ direction \cite{Maradudin}.  For the transverse modes, where the zone boundary occurs at $h$ = 0.33, there is no collision between the strongest modes, presumably due to the lower transverse sound velocity.

(ii) If the positions of the optic and acoustic branches are taken from the self-consistent phonon calculations and the existence of a third dispersive excitation is assumed, excellent fits to the data can be obtained using damped harmonic oscillator lineshapes convoluted with a Gaussian instrumental resolution function of 0.2 meV FWHM (Fig.~\ref{fig:intersect}).   The optic branch corresponds to the broad, weak excitation at higher energies, with a significantly broadened linewidth.  This assignment resolves the discrepancy between the self-consistent phonon calculations and experimental observations. Close to their intersection both the acoustic branch and the vacancy mode widen significantly, indicating decay of phonons into vacancy waves.  Above the intersection, just one sharp mode persists.  

Markovich {\it et al.}~\cite{MPBF02} observed an optic-like mode in the bcc phase of $^4$He in a longitudinal cut through the (110) Bragg peak, although none should exist in a monatomic bcc structure.  The strong similarity between this spectrum and that reported here in the hcp phase suggests that this branch of the spectrum in the bcc phase could be the vacancy excitation.

With our identification of the intermediate energy mode as a vacancy excitation, the minimum energy is identified as the vacancy activation energy, and is found to be 11 K.  This is consistent with values obtained by NMR \cite{MHMH74} and observation of activated thermal expansion above $\approx$ 10 K in x-ray \cite{FGS89} and neutron diffraction \cite{BGSH+07} studies.  The higher activation energy observed in the bcc phase \cite{MHMH74} is consistent with the higher melting temperature of the bcc crystal.  The dispersion of this vacancy mode is quadratic, and so an effective mass can be extracted.  Parameters for this quadratic dispersion are given in Table \ref{tab:tab2}.  The uncertainties quoted include only those from the fit and therefore may underestimate the total uncertainty.  This excitation was resolution limited and from  our highest resolution data with incident neutron wavelength 5.2 \AA~we can place an upper bound of 21 $\mu$eV on the linewidth.  The possibility of a dispersive mode associated with defects (specifically $^3$He impurities) in $^4$He has been discussed by Manousakis \cite{Manousakis07}, although we note that the amount of $^3$He in our sample is less than 0.0008\%. 

\begin{table*}[bth]
\begin{tabular}{|c|c|c|c|}
\hline
Mode & Effective mass & Energy gap & Vacancy minimum \\
 &  m$_{\mathrm{He}}$ & meV & \AA $^{-1}$ \\
\hline
Longitudinal & 0.107 $\pm$ 0.003 & 0.951 $\pm$ 0.004 & 1.964 $\pm$ 0.006 \\
Crystal 1 & & &\\
\hline
Longitudinal & 0.114 $\pm$ 0.005 & 1.002 $\pm$ 0.003 & 1.970 $\pm$ 0.006 \\
Crystal 2 & & & \\
\hline
Transverse & 0.144 $\pm$ 0.002 & 0.916 $\pm$ 0.002 & 1.970 $\pm$ 0.001 \\
Crystal 1 & & &\\
\hline
Roton & 0.094 $\pm$ 0.002 & 0.680 $\pm$ 0.001  & 2.054 $\pm$ 0.001 \\
Superfluid (24 bar) & & & \\
\hline
\end{tabular}
\caption{Characteristics of the vacancy mode in the solid and the roton mode in the superfluid.}
\label{tab:tab2}
\end{table*}

To summarize, we have observed a gapped zone-center excitation that is not a conventional optic phonon, because the longitudinal optic and acoustic modes cannot cross and annihilate on symmetry grounds, and because a similar mode was observed in the bcc phase \cite{MPBF02}, which has no optic mode.  

\begin{figure}[tbhp]
\begin{center}
\includegraphics[width=0.4\textwidth]{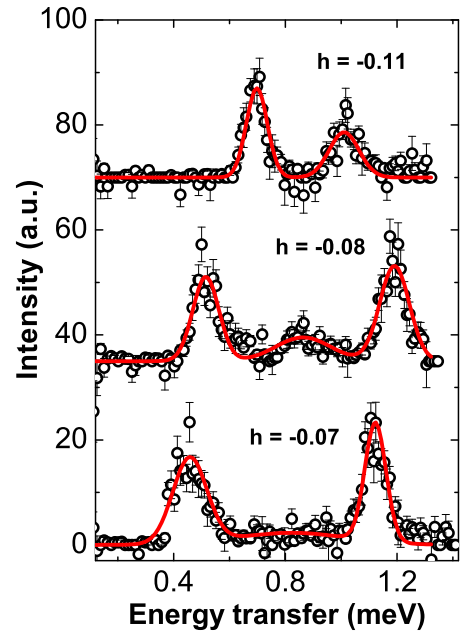}
\end{center}
\caption{Energy cuts at (1-$h$,2$h$,0) integrating over a momentum transfer of $h$ $\pm$ 0.005 r.l.u.~in the transverse direction and $\pm$ 0.01 r.l.u.~in the longitudinal direction, for three selected values of $h$.  The lines are Gaussian fits to the data.  The wavelength was 5.2 \AA, and these measurements were taken at 650 mK.  At $h$ = -0.11 the highest energy mode is out of the accessible energy range.  As $h$ decreases, the intensity of the middle mode decreases, finally disappearing at $h$ = -0.07.  The spectra are displaced vertically by 35 units.}
\label{fig:extra_mode}
\end{figure}

An additional mode is observed between the transverse acoustic and vacancy modes (Fig.~\ref{fig:extra_mode}).  This new mode possesses the lattice symmetry and therefore cannot belong to a second crystal, a possibility also dismissed by Laue scans of the crystal.  It is also observed at two different incident wavelengths, 5.2 and 3.7 \AA, indicating that it is not due to multiple scattering.  It is natural to associate both of these non-phonon features of the hcp $^4$He excitation spectrum with some form of propagating defects or vacancies.


In hcp $^4$He, the specific heat capacity is not fully explained by phonon contributions \cite{GHP73}, and it seems likely that this excess specific heat belongs, at least in part, to the delocalized vacancy excitations seen here.  Because of the substantial gap and small effective mass, thermal population of these modes does not account for an appreciable part of the excess specific heat.  It would appear that the T$^7$ term in the measured specific heat is due almost entirely to lattice anharmonicity.  This problem may ultimately be resolved by determination of the complete density of states for all excitations, whereas so far we have only measured the spectra for wavevector transfer in the basal plane.  

\section{Conclusions}

A new coherent excitation has been observed in single crystalline $^4$He.  We argue that it is the elementary excitation associated with delocalized vacancies in $^4$He and set a lower limit on its lifetime of 0.2 ns.  This new excitation, as interpreted here, makes an insignificant contribution to the excess specific heat at high temperatures.  We have also observed a broad feature in the spectrum which appears to be the optical phonon branch that was not observed in previous measurements.  The longitudinal acoustic phonon and vacancy modes intersect and merge into a single broadened mode, with a linear dispersion at high energies.

While to high accuracy there is no change in the density of states between 0.15 meV and 5.5 meV upon cooling through the putative supersolid transition, the new collective modes that we have observed remain an intriguing aspect of solid helium.  The similarities between the superfluid spectrum and the longitudinal part of the solid spectrum are remarkable and indicate that roton-like entities persist in the solid state, modified principally by the appearance of transverse dispersion.  

\section{Acknowledgements}
This work was supported in part by the National Science Foundation under Grant Nos.~DMR-0604295  and DMR-0454672.  The DAVE software package was used for elements of the data reduction and analysis (http://www.ncnr.nist.gov/dave).

\end{document}